\documentclass[conference]{IEEEtran}
\IEEEoverridecommandlockouts
\usepackage{cite}
\usepackage{amsmath,amssymb,amsfonts}
\usepackage{bbm}
\usepackage{algorithmic}
\usepackage{graphicx}
\usepackage{textcomp}
\usepackage{xcolor}
\usepackage{authblk}
\def\BibTeX{{\rm B\kern-.05em{\sc i\kern-.025em b}\kern-.08em
    T\kern-.1667em\lower.7ex\hbox{E}\kern-.125emX}}

\usepackage{tikz}
\usetikzlibrary{shapes,arrows}
\usepackage{adjustbox}
\usepackage[top=0.75in,bottom=1.05in,left=0.625in,right=0.625in]{geometry}

\begin{document}

\title{Communication via Sensing}

\author[*]{Mohammad Kazemi\thanks{M. Kazemi's work was funded by UK Research and Innovation (UKRI) under the UK government’s Horizon Europe funding guarantee [grant number 101103430]. T. M. Duman's work was funded by the European Union ERC TRANCIDS 101054904. D. G\"und\"uz acknowledges funding from UKRI under the ERC-Consolidator project AI-R (EP/X030806/1) and INFORMED-AI Hub (EP/Y028732/1).
Views and opinions expressed are however those of the author(s) only and do not necessarily reflect those of the European Union or the European Research Council Executive Agency. Neither the European Union nor the granting authority can be held responsible for them. 
}}
\author[**]{Tolga M. Duman}
\author[*]{Deniz G\"und\"uz}
\affil[*]{Dept. of Electrical and Electronic Engineering, Imperial College London  \protect\\ Email: \{mohammad.kazemi,d.gunduz\}@imperial.ac.uk}
\affil[**]{Dept. of Electrical and Electronics Engineering, Bilkent University
\protect\\Email:  duman@ee.bilkent.edu.tr}

\maketitle

\begin{abstract}
We present an alternative take on the recently popularized concept of `\textit{joint sensing and communications}', which focuses on using communication resources also for sensing. Here, we propose the opposite, where we utilize the receiver's sensing capabilities for communication. Our goal is to characterize the fundamental limits of communication over such a channel, which we call `{\it communication via sensing}'. We assume that changes in the sensed attributes, such as location and speed, are limited due to practical constraints, which are captured by assuming a finite-state channel (FSC) with an input cost constraint. We first formulate an upper bound on the \(N\)-letter capacity as a cost-constrained optimization problem over the input sequence distribution, and then convert it to an equivalent problem over the state sequence distribution. Moreover, by breaking a walk on the underlying Markov chain into a weighted sum of traversed graph cycles in the long walk limit, we obtain a compact single-letter formulation of the capacity upper bound. Finally, for a specific case of a two-state FSC with noisy sensing characterized by a binary symmetric channel (BSC), we obtain a closed-form expression for the capacity upper bound. Comparison with an existing numerical lower bound shows that our proposed upper bound is very tight for all crossover probabilities.
\end{abstract}


\section{Introduction}
In recent years, there has been significant interest in joint communication and sensing \cite{Liu22, Ahmadipour24}. The motivation is to repurpose existing communication devices and signals for sensing, e.g., the location or velocity of objects within the coverage area. Here, we formulate another related yet novel problem by asking the following question: Can we repurpose sensing devices, such as radar, LiDAR, or cameras, for communicating information? That is, is it possible to communicate by sensing?

\begin{figure}[t]
    \centering
    \includegraphics[width=0.9\columnwidth]{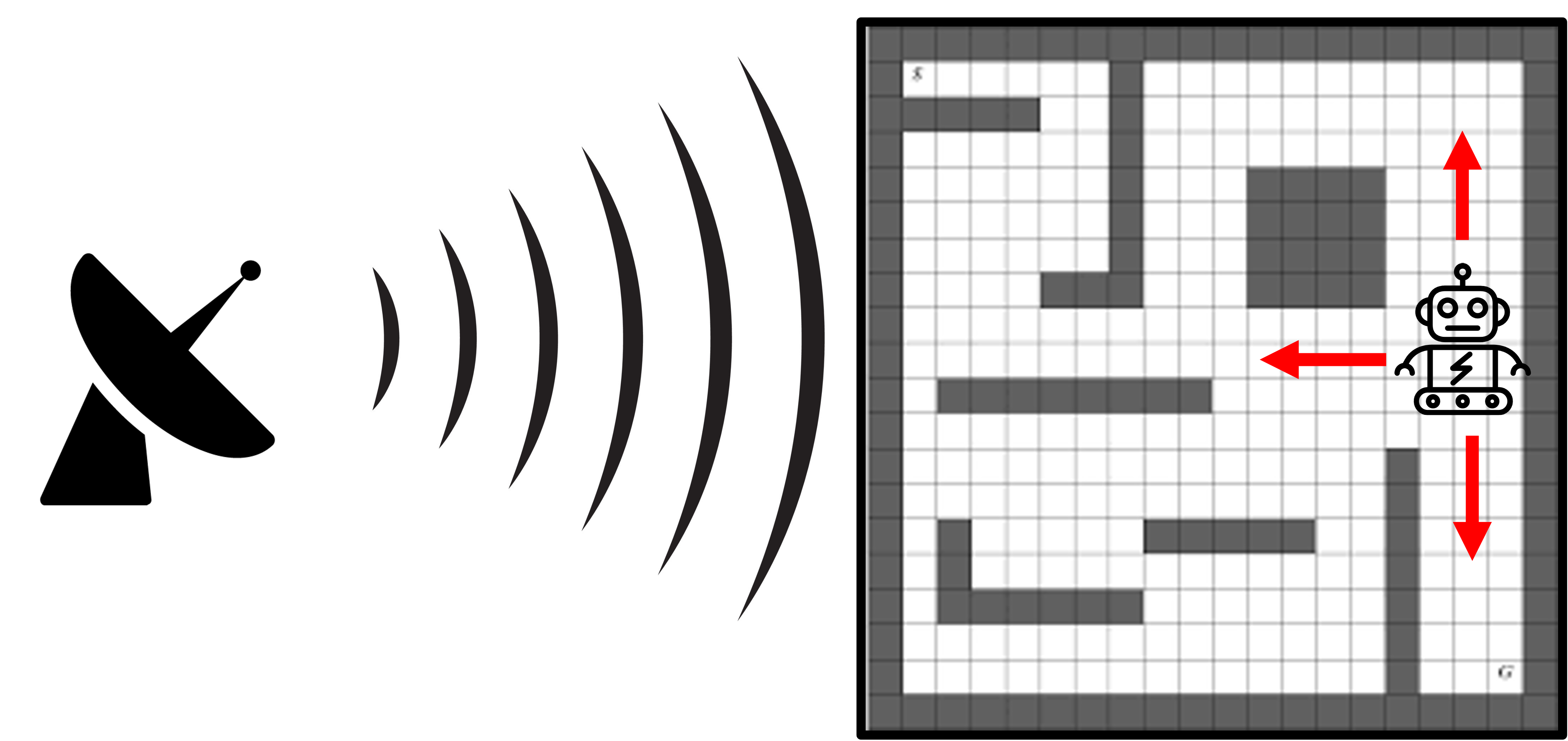}
    \caption{An example of communication via sensing, where the robot encodes information into its location, which can be sensed by a remote radar within some fidelity.}\label{figure:model}
\end{figure}

In conventional wireless communication systems, transceivers are equipped with dedicated modulation and demodulation capabilities to convey data from one point to another.
But what if a device does not have transmission capability, e.g., due to cost constraints? Is it still possible to transfer data to a receiver with conventional sensing capabilities? 
Assuming that the receiver is capable of sensing and tracking some attributes (e.g., location, orientation, or temperature) of a device that wants to convey information, the device can encode its data by varying the state of this attribute, which can then be sensed and decoded by the receiver. We call this communication paradigm {\it communication via sensing} (see Fig. \ref{figure:model} for an illustration). 
However, depending on the nature of the sensed attribute, such a channel typically imposes constraints on how the state can be changed over time, resulting in a channel with input constraints.

For the case in which the receiver can perfectly sense the device's attribute at each time instant (i.e., the noiseless scenario), if there are no restrictions on the values this attribute can take (apart from being limited to a discrete finite set $\mathcal{S}$), the communication capacity, $C$, is trivially given by
$C = \log (|\mathcal{S}|)$,
where $|\mathcal{S}|$ denotes the number of states (alphabet size) that the sensed attribute can take.

If transitions between some states are prohibited (or not possible) (for example, in Fig. \ref{figure:model}, the robot can move only in three directions at its current location), the relation between states can be visualized via a directed graph. It is shown in \cite{Shannon48} that the capacity of a noiseless input-constrained system is given by
$	C = \log (\lambda_{\max}),$
where $\lambda_{\max}$ is the largest eigenvalue of the adjacency matrix of the corresponding directed graph, which can be shown to be real and positive using the modified Perron–Frobenius theorem. 
For more results on the capacity of input-constrained systems without channel noise, see \cite{Marcus01} and the references therein.
With the additional cost (such as power) constraint, the authors in \cite{Justesen84} calculate the capacity of input-constrained systems as well as the transition probabilities of the Markov process that achieves it. In \cite{Neuhoff96}, a method is presented to construct finite-state codes for input-constrained systems by extending the state-splitting algorithm.

The literature on input-constrained systems with a noisy channel is, however, limited. Numerically computed capacity bounds are obtained in \cite{Arnold06, Kavcic01} using the iterative sum-product algorithm. The authors of \cite{Li14} provide a more compact formulation of the capacity for the special case of an erasure channel. The capacity with feedback is formulated as a computable optimization problem in \cite{Sabag16}, which serves as an upper bound on the capacity without feedback. 
Using the dual form of the capacity optimization problem, \cite{Vontobel01} presents a single-letter characterization of an upper bound on the capacity of noisy input-constrained systems.   
Following this result, by employing the Karush–Kuhn–Tucker (KKT) optimality conditions, closed-form expressions for the upper bound for some specific channels are obtained in \cite{Thangaraj17}.

In this paper, we provide a single-letter characterization of an upper bound on the capacity with cost constraint for the proposed {\it communication via sensing} paradigm, i.e., assuming that the receiver observes a noisy version of the channel states given state transitions and input cost constraints. The rest of the paper is organized as follows. We describe the system model in Section II, and obtain a single-letter formulation for the capacity upper bound in Section III. We then derive closed-form expressions for the upper bound in an example scenario in Section IV and conclude the paper in Section V.

\section{System Model}
In the proposed {\it communication via sensing} paradigm, the transmitter encodes its message by adjusting some of its attributes, whose states can be sensed and tracked by the receiver at fixed discretized time instants. The receiver decodes the message from its noisy observations of the state.

We model this as an irreducible finite-state channel (FSC), which captures the limitations of state transitions, such as those imposed by practical constraints. For instance, if the attribute is the transmitter's location, its changes may be bounded due to the transmitter's speed or energy limitations.
An FSC can be characterized by a tuple $(\mathcal{X} \times \mathcal{S}, P_{Y,S^+|X,S},\mathcal{Y} \times \mathcal{S})$, where $\mathcal{X}$, $\mathcal{Y}$, and $\mathcal{S}$ are the input, output, and channel state alphabets, respectively. 
In our model, $s_i \in \mathcal{S}$ denotes the state of the attribute that is sensed at time $i$, $x_i \in \mathcal{X}$ is the action of the transmitter at time $i$, and $y_i \in \mathcal{Y}$ is the noisy observation of the state by the receiver at time $i$.

Let $p_{Y,S^+|X,S}$ denote the channel law that governs the channel transition, i.e., the probability that the receiver observes $Y$ and the transmitter's state transitions to the new state $S^+$, given that the transmitter takes action $X$ in state $S$. 
We assume that the initial state $s_0 \in \mathcal{S}$ is known to both the transmitter and the receiver. 
Note that this assumption does not change the capacity since, regardless of the initial state, the stationary distribution of an irreducible Markov chain can be approached in $o(N)$ time, where $o(\cdot)$ indicates the little-$o$ notation and $N$ is the length of the input sequence.
Moving forward, some probability indices have been dropped for notational simplicity.

In the considered setup, $X$, $S$, and $Y$ form a Markov chain, that is, $X\rightarrow S \rightarrow Y$, and state transitions are deterministic given the current state and action pair. Therefore, in the specific scenario considered here, the conditional probability can be written as  
\begin{align}
    p\left(y^N,s^N |x^N\right) &= p\left(s^N |x^N\right)p\left(y^N|s^N,x^N\right) \nonumber\\
    &= p\left(s^N | x^N\right)p\left(y^N| s^N\right) \nonumber\\
        &= \mathbf{1}_{f(x^N)}(s^N) p\left(y^N | s^N\right), \label{eq15}
\end{align}
where $f(x^N)$ is the singleton containing the state sequence associated with the input sequence $x^N$ given the initial state $s_0$, and $\mathbf{1}_{A}(x)$ denotes the indicator function, i.e., is equal to one if $x\in A$ and zero otherwise.

We assume that actions in each state have an associated cost (e.g., required power), and that we are restricted to an average cost constraint. 
More specifically, we consider the average cost constraint $(k,\Gamma)$, i.e., $\frac{1}{N}\sum_{x^N} p(x^N) \sum_{i=1}^N k(x_i|s_{i-1}) \le \Gamma$, where $\Gamma$ is the maximum average cost and $\sum_{i=1}^N k(x_i|s_{i-1})$ represents the cost of employing the input sequence $x^N$ with $k(x_i|s_{i-1})$ being the cost of taking action $x_i$ in state $s_{i-1}$.

\section{Upper Bound on the Capacity}
In this section, we present a single-letter upper bound on the capacity of the sensing channel introduced above.
\subsection{Problem Formulation}
For a given initial state $s_0$, the $N$-letter channel capacity can be written as 
\begin{align}
    C_N = \max_{p(x^N) \in \mathcal{P}^N_X(\Gamma)} I(X^N;Y^N|S_0=s_0) , \label{cap-1}
\end{align}
where $\mathcal{P}^N_X(\Gamma)$ is the set of all input distributions on inputs of length $N$ that satisfy the average cost constraint.

The mutual information in \eqref{cap-1} can be expanded as
\begin{align}
    &I(X^N;Y^N|S_0=s_0) \nonumber\\
    &= H(Y^N|S_0=s_0) - H(Y^N|X^N, S_0=s_0)\\
     &= H(Y^N|S_0=s_0) - H(Y^N|X^N, S^N, S_0=s_0) \label{I0}\\
     &= H(Y^N|S_0=s_0) - H(Y^N|S^N, S_0=s_0) \label{I1}\\
     &= I(Y^N;S^N|S_0=s_0) \label{I2},
\end{align}
where in obtaining \eqref{I0} and \eqref{I1}, we used the fact that state transition is deterministic given the action and the previous state, and that $X^N$, $S^N$, and $Y^N$ form a Markov chain, i.e., $X^N \rightarrow S^N \rightarrow Y^N$, respectively.

Now, the capacity optimization over the input distribution is converted to one over the state distribution, constrained to the sequence of states limited by the underlying Markov chain. 
This is equivalent to the capacity formulation for an input-constrained system with $s^N$ as input, $y^N$ as output, and $\mathcal{S}^N$ as the constrained set of possible inputs. Therefore, combining \eqref{I2} and \eqref{cap-1}, the channel capacity can be written equivalently as
\begin{align}
    C_N = \max_{p(s^N) \in \mathcal{P}^N_S(\Gamma)} I(S^N;Y^N|S_0=s_0) , \label{cap-2}
\end{align}
where $\mathcal{P}^N_S(\Gamma)$ is the set of all distributions of state sequences of length $N$ that satisfy the average cost constraint; i.e., $\frac{1}{N}\sum_{s^N\in \mathcal{S}^N} p(s^N) \sum_{i=1}^N k(s_i|s_{i-1}) \le \Gamma$.
Moving forward, to simplify the analysis, we work with a simplified (pruned) version of the underlying Markov chain, where every pair of states is connected by at most one action; that is, from any state, no two actions lead to the same next state. Note that in the case without cost constraints, there is no priority in which edge (branch) to keep. However, in the cost-constrained case, the edge with the lowest cost is kept. In this case, the cost of transition from state $s_{i-1}$ to $s_i$ is defined as $k(s_i|s_{i-1}):=\min_{\substack{x_i\in \mathcal{X}:\\p(s_i|s_{i-1},x_i)=1}}k(x_i|s_{i-1})$.
An example of this pruning process is depicted in Fig. \ref{equiv} for an FSC with $\mathcal{X}=\{x^1,x^2\}$, where the taken actions and their associated costs are denoted by labels with $k(s^j|s^i)$ being the cost of transition from state $s^i$ to $s^j$.

\subsection{$N$-Letter Capacity Upper Bound}
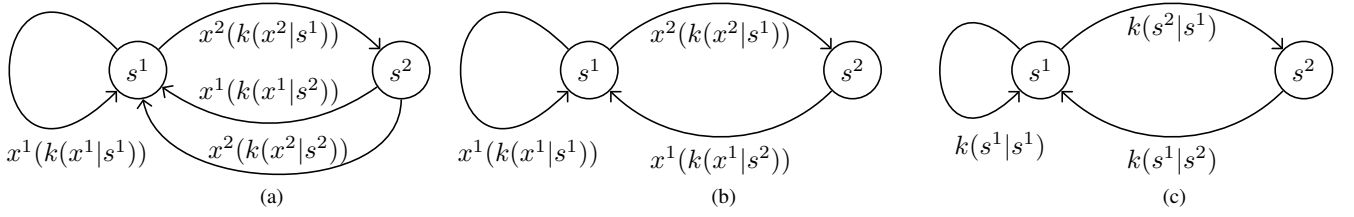
\begin{figure*}[htb]
  \centering
  \tikzstyle{cnode}=[circle,draw]
  \begin{adjustbox}{trim=0 .5cm 0 1.3cm}
  \begin{tikzpicture}
    \begin{scope}[shift={(-2,0)},node distance=2.5cm,>=angle 90,semithick]
        \node[cnode] (n1)                   {$s^1$};
        \node[cnode] (n2) [right of=n1,xshift=1cm]     {$s^2$};
        \draw[->] (n1) edge[loop left,out=135,in=225,min distance=2.7cm] node[below,xshift=.9cm,yshift=-.8cm]{$x^1 (k(x^1|s^1))$} (n1);
        \draw[->] (n1) to[out=45,in=135] node[below,yshift=-.1cm]{$x^2 (k(x^2|s^1))$} (n2);
        \draw[->] (n2) to[out=215,in=325] node[above,yshift=.1cm]{$x^1 (k(x^1|s^2))$} (n1);
        \draw[->] (n2) to[out=270,in=280]  node[above]{$x^2 (k(x^2|s^2))$} (n1);
    \end{scope}
    \node[anchor=west] at (-.5,-1.7) {\footnotesize (a)};
    \begin{scope}[shift={(4,0)},node distance=2.5cm,>=angle 90,semithick]
        \node[cnode] (n1)                   {$s^1$};
        \node[cnode] (n2) [right of=n1,xshift=1cm]     {$s^2$};
        \draw[->] (n1) edge[loop left,out=135,in=225,min distance=2.7cm] node[below,xshift=.9cm,yshift=-.8cm]{$x^1 (k(x^1|s^1))$} (n1);
        \draw[->] (n1) to[out=45,in=135]  node[below,yshift=-.1cm]{$x^2 (k(x^2|s^1))$} (n2);
        \draw[->] (n2) to[out=225,in=315]   node[below]{$x^1 (k(x^1|s^2))$} (n1);
    \end{scope}
    \node[anchor=west] at (5.5,-1.7) {\footnotesize (b)};
    \begin{scope}[shift={(10,0)},node distance=2.5cm,>=angle 90,semithick]
        \node[cnode] (n1)                   {$s^1$};
        \node[cnode] (n2) [right of=n1,xshift=1cm]     {$s^2$};
        \draw[->] (n1) edge[loop left,out=135,in=225,min distance=2cm] node[below,xshift=.8cm,yshift=-.7cm]{$k(s^1|s^1)$} (n1);
        \draw[->] (n1) to[out=45,in=135]  node[below]{$k(s^2|s^1)$} (n2);
        \draw[->] (n2) to[out=225,in=315]   node[below]{$k(s^1|s^2)$} (n1);
    \end{scope}
    \node[anchor=west] at (11.5,-1.7) {\footnotesize (c)};
  \end{tikzpicture} 
  \end{adjustbox}
\caption{The Markov chains governing (a) the FSC, (b) its pruned version (assuming $k(x^1|s^2)\le k(x^2|s^2)$), and (c) its equivalent input-constrained model.}
\label{equiv}
\end{figure*}

A state sequence can be decomposed into a combination of a path and some cycles on the underlying Markov chain. For example, in the FSC in Fig. \ref{equiv} (c), the state sequence $s^1s^1s^2s^1s^2$ can be decomposed into two cycles, $s^1s^1$ and $s^1s^2s^1$, and a path $s^1s^2$.
In the remainder of this section, we first formulate an upper bound on the $N$-letter capacity. Then, by using the state sequence decomposition approach, we obtain a single-letter formulation for the capacity upper bound in the form of a calculable compact optimization problem.

Compensation identity \cite[Theorem 9.1]{Topsoe01} states that for any test distribution on the channel output, $q_{Y^N}(\cdot)$, (not necessarily due to the input distribution $p_{S^N}(\cdot)$), we have
\begin{align}
    I(S^N;Y^N) =& \sum_{s^N\in \mathcal{S}^N} p_{S^N}(S^N) D\left(p_{Y^N|S^N}(\cdot|s^N)\|q_{Y^N}(\cdot)\right)  \nonumber\\
    &- D\left(p_{Y^N}(\cdot)\|q_{Y^N}(\cdot)\right),
\end{align}
where $D(\cdot ||\cdot)$ indicates the Kullback–Leibler (KL) divergence (relative entropy).

This gives us an upper bound on the capacity as
\begin{align}
        C_N \!\le \!\max_{p(s^N):s^N\in \mathcal{S}^N}  \!\!\sum_{s^N\in \mathcal{S}^N} \!\!p_{S^N}(s^N) D\!\left(  p_{Y^{\!N}|S^{\!N}}\!\left(\cdot | s^N\right) \!\|q_{Y^{\!N}}(\cdot)\!\right) \!, \label{opt1}
\end{align}
where equality holds when the chosen test distribution is the output distribution corresponding to the optimal input distribution (which means $D\left(p_{Y^N}(\cdot)\|q_{Y^N}(\cdot)\right)=0$).

Since the objective function in the optimization problem \eqref{opt1} is linear in $p(s^N)$, for the unconstrained case, it is maximized when all the weight is dedicated to $s^N$ with the maximum coefficient.
This gives us the following capacity upper bound, called the \textit{dual capacity} upper bound \cite{Csiszar11}:
\begin{align}
    C_{UB} = \lim_{N \rightarrow \infty} \frac{1}{N} \max_{s^N\in \mathcal{S}^N} D\left(p_{Y^{N}|S^{N}}(\cdot|s^N)\|q_{Y^N}(\cdot)\right). \label{dual} 
\end{align}
This shifts the optimization over the state distribution to a search over valid state sequences, which simplifies the optimization.
Note that the dual capacity upper bound still holds in the cost-constrained case; however, it is not very tight since the sequence that maximizes $D\left(p_{Y^{N}|S^{N}}(\cdot|s^N)\|q_{Y^N}(\cdot)\right)$ may not satisfy the cost constraint. In the following, we provide a tighter upper bound for the cost-constrained case.

In this paper, we consider the average cost constraint $(k,\Gamma)$, 
which, for the simplified underlying Markov chain, can be written as $\frac{1}{N}\sum_{s^N\in \mathcal{S}^N} p(s^N) \sum_{i=1}^N k(s_i|s_{i-1}) \le \Gamma$. So, the capacity upper bound for the cost-constrained case can be written as
\begin{align}
   & C_{UB} = \nonumber\\
   & \!\!\lim_{N \rightarrow \infty} \!\frac{1}{N} \!\!\max_{\substack{p(s^N)\\ \in \mathcal{P}^N_S(\Gamma) }}  \sum_{s^N\in \mathcal{S}^N} \!\!\!p(s^N) D\!\left(  p_{Y^{N}|S^{N}}\left(\cdot | s^N\right) \|q_{Y^N}(\cdot)\right)\!. \label{cap2}
\end{align}

This bound is not easily calculable since it depends on the input length $N$. In the following, we devise a technique to convert it to an easily calculable single-letter formulation.

\subsection{Single-Letter Capacity Upper Bound}
To obtain a computable expression, we take the test distribution to be a Markov process with unit memory; i.e., $q(y_n | y^{n-1})=q(y_n | y_{n-1})$.

With this assumption, using the result in \cite{Vontobel01}, the KL divergence in \eqref{opt1} can be reformulated as 
\begin{equation}  
	D\left(p_{Y^{N}|S^{N}}{(\cdot|s^N)} || q_{Y^N}{(\cdot)}\right) = \sum_{n=1}^N m_{q,p}(s_{n-1},s_n),
	\label{KL}
\end{equation}
where 
\begin{align}  
    &m_{q,p}(s_{n-1},s_n):= \sum_{y \in \mathcal{Y}} p_{Y|S}{(y|s_{n-1})}  D\left(p_{Y|S}{(\cdot|s_n)} || q_{Y}{(\cdot|y)}\right) \nonumber\\
    &=\!\sum_{y_1,y_2\in\mathcal{Y}} \!\!\!p_{Y|S}(y_1|s_{n-1}) p_{Y|S}(y_2|s_n) \log \frac{p_{Y|S}(y_2|s_n)}{q_{Y}(y_2|y_1)} \label{metric}
\end{align}
is the metric corresponding to the transition from state $s_{i-1}$ to $s_i$ given the output test distribution $q_{Y}$.

Plugging this into \eqref{opt1}, the capacity upper bound becomes
\begin{equation}
    C_{UB} \!= \!\lim_{N \rightarrow \infty} \frac{1}{N} \!\max_{\substack{p(s^N)\in \mathcal{P}^N_S(\Gamma) }}  \sum_{s^N\in \mathcal{S}^N} \!\!\!p(s^N) \sum_{n=1}^N m_{q,p}(s_{n-1},s_n).
	\label{dual_UB2}
\end{equation}

Any state sequence represents a \textit{walk} on the underlying Markov chain, which can be decomposed into several cycles and a direct path, which is the shortest path that connects the first and last points of the walk. It is easy to show that the contribution of the shortest path to the walk metric (sum of the metrics of the state transitions on the walk) is $o(N)$, so only cycle metrics matter in the limit. Therefore, for the objective function in \eqref{dual_UB2}, we have
\begin{equation}
	\begin{split}
	&\lim_{N\rightarrow \infty} \frac{1}{N}  \sum_{s^N\in \mathcal{S}^N} \!\!p(s^N)\sum_{n=1}^N m_{q,p}(s_{n-1},s_n)\\
	&= \lim_{N\rightarrow \infty} \frac{1}{N} \sum_{s^N\in \mathcal{S}^N} \!\!p(s^N)\sum_{i=1}^{N_c} n_i(s^N) m_{q,p}(c_i) \\
	&= \sum_{i=1}^{N_c} \mu_i \bar m_{q,p}(c_i),
	\end{split}
	\label{dual_UB3}
\end{equation}
where $\bar m_{q,p}(c_i) := \frac{ m_{q,p}(c_i)}{l(c_i)}$ is the normalized metric of the $i$-th distinct cycle $c_i$ of the pruned Markov chain with length $l(c_i)$, 
$N_c$ is the total number of distinct cycles, $n_i(s^N)$ is the number of times the $i$-th distinct cycle is visited on the walk associated with $s^N$, and $\mu_i:=\lim_{N\rightarrow \infty} \frac{1}{N} \sum_{s^N\in \mathcal{S}^N} \!p(s^N) n_i(s^N) l(c_i)$ is the average frequency of state transitions in the $i$-th distinct cycle being visited over all valid state sequences. As an example, in the FSC in Fig. \ref{equiv} (c), there are two cycles (hence, $N_c=2$): a loop $c_1=(s^1s^1)$ and a cycle $c_2=(s^1s^2s^1)$, with lengths $l(c_1)=1$ and $l(c_2)=2$, and average metrics $\bar m_{q,p}(c_1) =m_{q,p}(s^1,s^1)$ and $\bar m_{q,p}(c_2) = 0.5(m_{q,p}(s^1,s^2)+m_{q,p}(s^2,s^1))$, respectively. For a sample state sequence $s^1s^1s^1s^1s^2s^1$ (hence, $N=5$), we have $n_1(s^N)=3$, $n_2(s^N)=1$.

Knowing that $l(s^N)=N$ and that the length of the shortest path is $o(N)$, the sequence length can be expanded as follows
\begin{equation}
	\begin{split}
	\frac{1}{N} l(s^N)\!
    &=\lim_{N\rightarrow \infty} \!\frac{1}{N} \sum_{s^N\in \mathcal{S}^N} \!p(s^N) l(s^N)\\
	&= \lim_{N\rightarrow \infty} \!\!\frac{1}{N} \sum_{s^N\in \mathcal{S}^N} \!\!p(s^N) \sum_{i=1}^{N_c} n_i(s^N) l(c_i) \\
	&= \sum_{i=1}^{N_c} \mu_i.
	\end{split}
	\label{dual_UB3.5}
\end{equation}

Similarly, we can write the cost constraint as a function of the average costs of the cycles as follows.
\begin{equation}
	\begin{split}
	&\lim_{N\rightarrow \infty} \frac{1}{N} \sum_{s^N\in \mathcal{S}^N} p(s^N) \sum_{i=1}^N k(s_i|s_{i-1}) \\
	&= \lim_{N\rightarrow \infty} \!\frac{1}{N} \!\!\sum_{s^N\in \mathcal{S}^N} \!\!p(s^N) \!\!\left( \sum_{i=1}^{N_c} n_i(s^N) l(c_i) \bar P(c_i) \!+\! o(N) \!\right) \\
	&= \sum_{i=1}^{N_c} \mu_i \bar P(c_i),
	\end{split}
	\label{dual_UB3.75}
\end{equation}
where $\bar P(c_i)$ is the average cost of cycle $c_i$.

We observe that both the objective function and the cost constraint can be represented as functions of $\mu_i$. Finally, combining \eqref{dual_UB2}-\eqref{dual_UB3.75} and knowing that the upper bound is valid for any test distribution, a single-letter capacity upper bound for the cost-constrained case can be obtained by solving
\begin{equation}
	\begin{split}
		C_{UB} = \min_{q \in \mathcal{Q}} &\max_{0\le\mu_i \le 1} \sum_{i=1}^{N_c} \mu_i\bar m_{q,p}(c_i),\\
            s.t. & \hspace{4pt} \sum_{i=1}^{N_c} \mu_i=1,\\
            &\sum_{i=1}^{N_c} \mu_i \bar P(c_i) \le \Gamma.
	\end{split}
	\label{dual_UB4}
\end{equation}
where $\mathcal{Q}$ is the set of all possible test distributions.

It is easy to show that for the case without a cost constraint, the solution to the optimization problem in \eqref{dual_UB4} is to allocate the entire weight to the cycle with the maximum normalized metric. This gives us the maximum normalized cycle metric as an upper bound on the capacity, as obtained in \cite{Thangaraj17}.

\section{An Example: Binary Symmetric Channel }
To further study the proposed single-letter upper bound on the capacity of the cost-constrained case, we focus on a specific setup, namely, the FSC depicted in Fig. \ref{equiv} with noisy sensing characterized by a BSC with crossover probability $p$; i.e., $p_{Y|S}=\begin{bmatrix}1-p&p\\p&1-p\end{bmatrix}$.
  
For the output, we consider the non-symmetric Markov test distribution $q_Y(y_2|y_1)$ defined as 
$q_Y(y_2|y_1)=\begin{bmatrix}
a&1-a\\
1-b&b
\end{bmatrix}$,
where $0< a,b <1$, and the rows and columns correspond to $y_1=0,1$ and $y_2=0,1$, respectively. 

Using \eqref{metric} and defining $\bar x:= 1-x$, for the pruned Markov chain and the BSC, the branch metrics become
\begin{align*}
  m_{q,p}(s^1,s^1)
  &= -H(p) - \bar p^2\log{a} -p\bar p\log{\bar a \bar b} -p^2\log{b},\\
  m_{q,p}(s^1,s^2)
  &= -H(p) - \bar p^2\log{\bar a} -p\bar p\log{a b} -p^2\log{\bar b},\\
  m_{q,p}(s^2,s^1)
  &= -H(p) - \bar p^2\log{\bar b} -p\bar p\log{a b} -p^2\log{\bar a},
\end{align*}

Examining the equivalent pruned model in Fig. \ref{equiv}, we have two cycles, 
with average metrics $\bar m(c_1) =m_{q,p}(s^1,s^1)$ and $\bar m(c_2) = 0.5(m_{q,p}(s^1,s^2)+m_{q,p}(s^2,s^1)) = -H(p) - 0.5 (p^2 +\bar  p^2)\log{\bar a\bar b} -p\bar p\log{a b}$, and average costs $\bar P(c_1)= k(s^1|s^1)$ and $\bar P(c_2)= 0.5(k(s^2|s^1)+k(s^1|s^2))$, respectively.
Plugging these into the optimization problem in \eqref{dual_UB4}, we have
\begin{align}
    C_{UB} &= \min_{0< a,b < 1} \max_{0 \le \mu \le 1} \mu \bar m(c_1)+(1-\mu) \bar m(c_2) \label{minmax}\\
    & \quad \quad s.t. \quad \quad \mu \bar P(c_1)+(1-\mu) \bar P(c_2) \le \Gamma \nonumber\\
    &= \max_{0 \le \mu \le 1} \min_{0< a,b< 1}  \mu \bar m(c_1)+(1-\mu) \bar m(c_2) \label{maxmin}\\
    & \quad \quad s.t. \quad \quad \mu \bar P(c_1)+(1-\mu) \bar P(c_2) \le \Gamma, \nonumber
\end{align}
where the second equality is due to the \textit{Sion's minimax theorem} \cite{Sion58} since it can be shown that given $\mu$, the objective function is a convex function of $a$ and $b$, and given $a$ and $b$, it is a concave (more specifically, linear) function of $\mu$ with a convex (more specifically, linear) constraint. 

Given $\mu$, since the constraint is independent of $a$ and $b$, for the minimization, we have
\begin{align}
    C_{min}(\mu)&:= \min_{0< a,b < 1} \mu \bar m(c_1)+(1-\mu ) \bar m(c_2).
\end{align}

As can be seen, it is a separable optimization problem, i.e.,
\begin{align}
    C_{min}(\mu)&= -H(p) + \min_{0< a < 1} g(a;\mu) + \min_{0< b < 1} h(b;\mu), \label{Cmin}
\end{align}
where
    $g(a;\mu) 
        : -\bar p (\bar p\mu + p \bar \mu) \log{a} - (\mu p\bar p +0.5\bar \mu (p^2 +\bar  p^2)) \log{\bar a}$ and
    $h(b;\mu) 
    := -p (p\mu + \bar p \bar \mu) \log{b} - (\mu p\bar p +0.5\bar \mu (p^2 +\bar  p^2)) \log{\bar b}$.

It is easy to show that if $\alpha_1,\alpha_2\le0$, $f(x)=\alpha_1\log x + \alpha_2 \log \bar x$, is a convex function of $x$ in the interval $(0,1)$ with the optimal point $x^*=\frac{\alpha_1}{\alpha_1+\alpha_2}$, which lies within the interval. 
Applying this to the minimization subproblems in \eqref{Cmin}, we get
\begin{align}
    a^*
    =\frac{-\bar p(p-\bar p)\mu+p \bar p}{(0.5-p)\mu+0.5} , \quad
    b^*
    =\frac{ p(p-\bar p)\mu+p \bar p}{-(0.5-p)\mu+0.5}.
\end{align}

Now, if we look at the original min-max problem in \eqref{minmax}, for the inner optimization, we have:
 \begin{align}
    C_{max} &:=\max_{0 \le \mu \le 1} \mu \bar m(c_1)+(1-\mu) \bar m(c_2) \\
    & \quad \quad s.t. \quad \mu \bar P(c_1)+(1-\mu) \bar P(c_2) \le \Gamma, \nonumber
\end{align}
which is a linear program. If $\bar P(c_2) \ge \bar P(c_1)$, combined with $0 \le \mu \le 1$, the constraint is simplified to $\mu_{th}\le \mu \le1$, where $\mu_{th}=\frac{\bar P(c_2) - \Gamma}{\bar P(c_2) - \bar P(c_1)}$. Now, if $\bar m(c_1) \ge \bar m(c_2)$, the objective function is maximized when $\mu=1$, otherwise, when $\mu = \frac{\bar P(c_2) - \Gamma}{\bar P(c_2) - \bar P(c_1)}$. The results for the $\bar P(c_2) \le \bar P(c_1)$ case can be obtained similarly. That is, if $\bar m(c_1) \ge \bar m(c_2)$, the objective function is maximized when $\mu=\mu_{th}$, otherwise, when $\mu = 0$.  So, we observe that the optimal $\mu$ always lies in its boundaries, and we only need to find the boundary with the higher value. Finally, combining this with the obtained results for the equivalent max-min problem, we have
\begin{align}
   &C_{UB}
        = -H(p) \\
        &-
        \begin{cases}
        \min\{gh(a^*,b^*;1),gh(a^*,b^*;\mu_{th})\},& \text{if} \bar P(c_2) \ge \bar P(c_1) ,\\
        \min\{gh(a^*,b^*;0),gh(a^*,b^*;\mu_{th})\},& \text{otherwise} ,
        \end{cases} \nonumber
\end{align}
where $gh(a,b;\mu):=g(a;\mu) + h(b;\mu)$.

\begin{figure}[t]
    \centering
    \includegraphics[trim={1cm 0.3cm 1.5cm 0.5cm},clip,width=.9\linewidth]{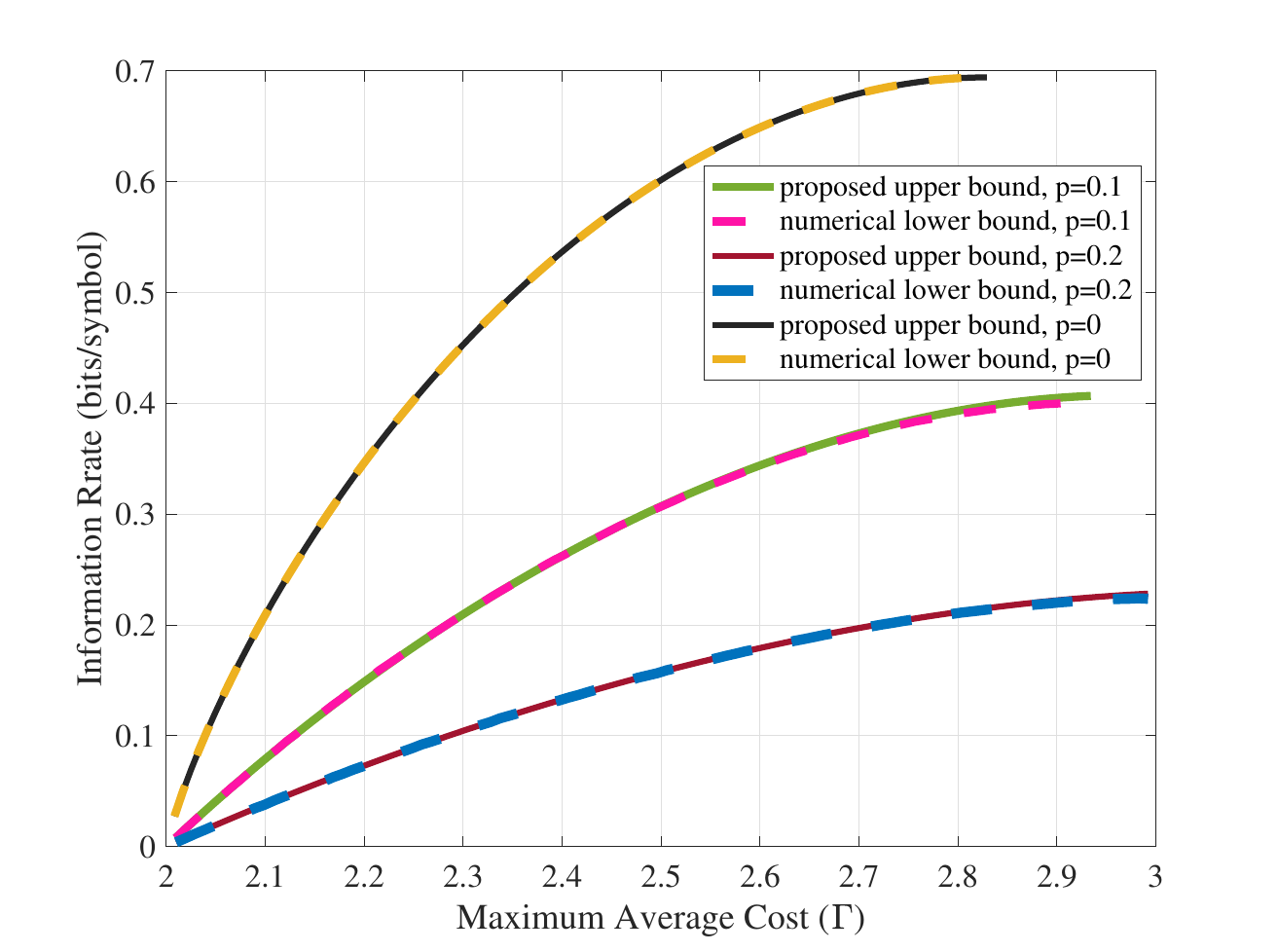}
    \caption{Information rate versus the maximum average cost.}\label{result}
\end{figure}
For $k(s^1|s^1)=2$, $k(s^2|s^1)=3$, and $k(s^2|s^1)=4$, the proposed upper bound is presented in Fig. \ref{result} as a function of the maximum average cost $\Gamma$ for different values of the crossover probability $p$. We also present a lower bound by modifying the unconstrained lower bound in \cite{Kavcic01}, based on the iterative sum-product algorithm, to account for the cost constraint by starting the iteration from a feasible point and projecting the output of each iteration to the feasible set, where feasibility means satisfying the maximum average cost $\Gamma$. We observe that the upper and lower bounds are very tight for the whole range of maximum average cost $\Gamma$ and the different values of crossover probability $p$.
Note that we plot each diagram in the interval $\Gamma_{\min} \le \Gamma \le \Gamma_{\max}$, where $\Gamma_{\min}$ is the minimum average cost dictated by the lowest average cost among the cycles of the underlying simplified (pruned) Markov chain, i.e., $\Gamma_{\min}=\min_{i\in [N_c]} \bar P(c_i)$. For $\Gamma \le \Gamma_{\min}$, the system capacity is zero, as the cost constraint can never be satisfied. On the other hand, $\Gamma_{\max}$ is equal to the average cost associated with the capacity-achieving input distribution for the case without cost constraint, which, unlike $\Gamma_{\min}$, depends on the channel parameters (in this example, the crossover probability $p$). For $\Gamma \ge \Gamma_{\max}$, the cost constraint is ineffective, and the capacity remains constant.

\section{Conclusions}
We introduced a new communication paradigm called {\it communication via sensing}. In this paradigm, a transmitter that is not equipped with conventional communication equipments encodes its data by varying some of its attributes, such as its location or speed that can be sensed and tracked by the receiver. The receiver can then decode the data using the observed states of the attributes. 
Moreover, we assume that there are some practical limitations in varying these attributes, such as state transition and cost constraints, under which we obtain a single-letter characterization of an upper bound on the capacity.
Finally, as an illustrative example, we consider an FSC with noisy observations characterized by a BSC channel, and show that the proposed upper bound is very tight for all crossover probabilities.


\end{document}